\documentclass[12pt]{iopart}
\expandafter\let\csname equation*\endcsname\relax
\expandafter\let\csname endequation*\endcsname\relax

\usepackage[utf8x]{inputenc}
\usepackage{cite}
\usepackage{bbm} 
\usepackage{amsfonts,amsmath,amssymb,stmaryrd}
\usepackage{amssymb}
\usepackage{mathtools}
\usepackage{braket}
\usepackage{graphicx}
\usepackage{hyperref}
\usepackage[capitalise]{cleveref}
\usepackage{mathrsfs}
\usepackage{verbatim}

\usepackage[dvipsnames]{xcolor}
\usepackage{soul}
\usepackage[T1]{fontenc}
\usepackage[caption=false]{subfig}

\usepackage{orcidlink}

\newcommand{\hH}{\hat{H}}
\newcommand{\hr}{\hat{r}}
\newcommand{\hp}{\hat{p}}
\newcommand{\hPB}{\hat{P}_\text{B}}
\newcommand{\pB}{P_\text{B}}
\newcommand{\hXB}{\hat{X}_\text{B}}
\newcommand{\gib}{g_\text{IB}}

\newcommand{\dd}{\text{d}}
\newcommand{\hphi}[1]{\hat{\phi}(#1)}
\newcommand{\hphid}[1]{\hat{\phi}^\dagger(#1)}
\newcommand{\half}{\tfrac{1}{2}}
\newcommand{\LLP}{\text{LLP}}
\newcommand{\xit}{\tilde{\xi}}

\begin{document}
\title{Dynamics of polaron formation in 1D Bose gases in the strong-coupling regime}

\author{Martin Will \orcidlink{0000-0003-1490-8928} and Michael Fleischhauer\orcidlink{0000-0003-4059-7289}}
\address{Department of Physics and Research Center OPTIMAS, University of Kaiserslautern-Landau, 67663 Kaiserslautern, Germany}

\vspace{10pt}
\begin{indented}
\item[]
\end{indented}


\begin{abstract}
We discuss the dynamics of the formation of a Bose polaron when an impurity is injected into a weakly interacting one-dimensional Bose condensate. 
While for small impurity-boson couplings this process can be described within the Froehlich model as generation, emission and binding of Bogoliubov phonons, this is no longer adequate if the coupling becomes strong. 
To treat this regime we consider a mean-field approach beyond the Froehlich model which accounts for the backaction to the condensate, complemented with Truncated Wigner simulations to include quantum fluctuation.
For the stationary polaron we find a periodic energy-momentum relation and non-monotonous relation between impurity velocity and polaron momentum including regions of negative impurity velocity.
Studying the polaron formation after turning on the impurity-boson coupling quasi-adiabatically and in a sudden quench, we find a very rich scenario of dynamical regimes. Due to the build-up of an effective mass, the impurity is slowed down even if its initial velocity is below the Landau critical value. For larger initial velocities we find deceleration and even backscattering caused by emission of density waves or grey solitons and subsequent formation of stationary polaron states in different momentum sectors.  In order to analyze the effect of quantum fluctuations we consider a trapped condensate to avoid 1D infrared divergencies. Using Truncated Wigner simulations in this case we show under what conditions the influence of quantum fluctuations is small.
\end{abstract}

\date{\today}
\maketitle


\section{Introduction}

The dynamics of a quantum impurity coupled to an interacting many-body environment is one of the most fundamental problems of many-body physics. Of particular interest is the dressing of the impurity with elementary excitations of the host systems leading to the formation of a quasiparticle. A  paradigmatic model of such a quasiparticle in condensed matter physics is the polaron, introduced by Landau and Pekar \cite{Landau1933,Pekar1946} to describe
the interaction of an electron with lattice vibrations 
in a solid, and which is key for understanding transport, response and induced interactions in many systems. In recent years ultra-cold quantum gases have become a versatile experimental testing ground for studying polaron physics with high precision and in novel regimes. For example, employing Feshbach resonances \cite{Chin2010} for neutral atoms, the impurity-bath interaction can be tuned from weak to strong coupling. Furthermore
  many-body environments of different quantum statistics and with different interactions can be considered. While impurities in a degenerate Fermi gas, called Fermi-polarons, have been studied in a number of experiments  
  only a small number of experiments exist on Bose polarons 
\cite{Catani2012,Jorgensen,Hu,Yan2019}. Here due to the large
 compressibility of the Bose gas a larger amount of excitations can be created by the impurity and interactions among the environment particles become increasingly important.

\begin{figure}[ht]
    \centering
     \includegraphics[width=0.6\textwidth]{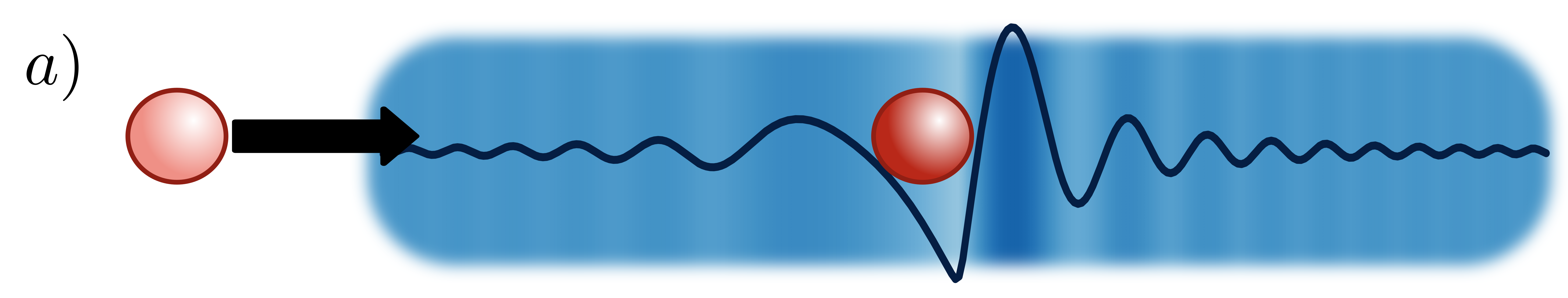}
     \includegraphics[width=\textwidth]{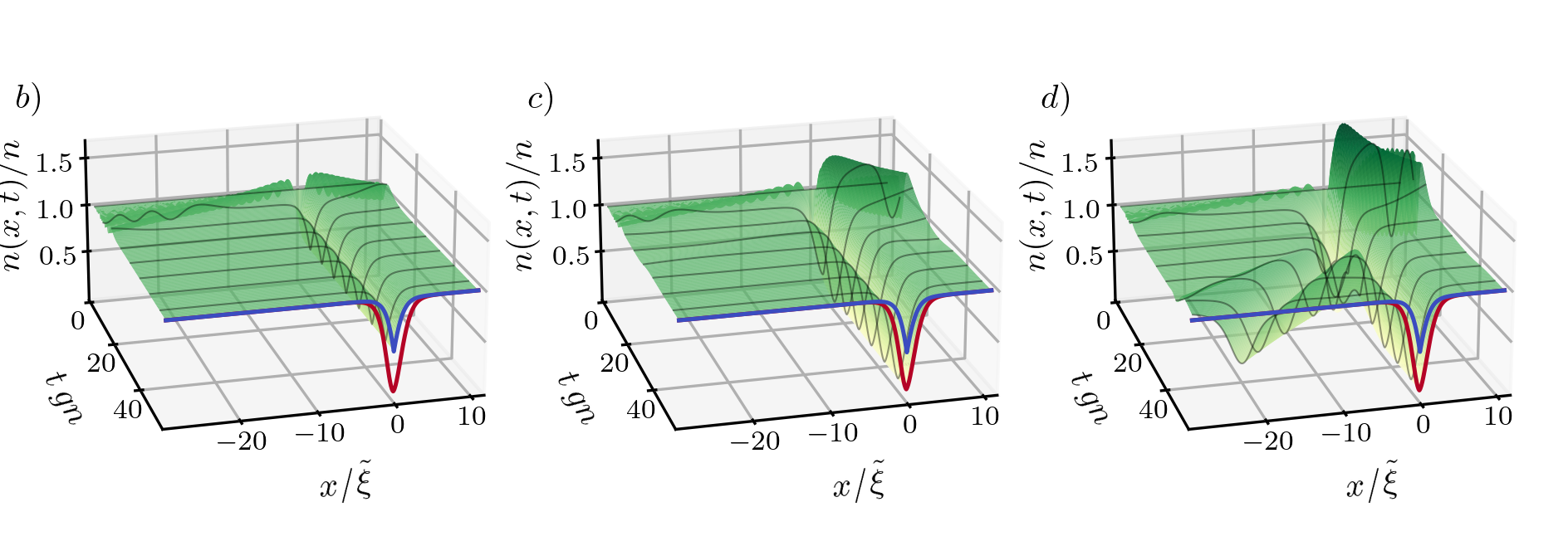}
    \caption{Formation of a polaron when an impurity is injected into a weakly interacting Bose gas (a). b)-d) 
    Evolution of the Bose gas density $n(x,t)$ for different total momenta ($p/M\tilde c=0.3, 1.0, 1.6$), where $M$ is the impurity mass, $n$ the average boson density, $\tilde c$ the speed of sound and $\tilde \xi$ the healing length (see text). The impurity emits density waves before converging into on of two stationary states, marked in red and blue. For large momenta (d) also solitons are created which can lead to a change in the direction of motion of the impurity. The evolution is shown for an impurity-Bose mass ratio $M = 10 m$, Tonks parameter $\gamma = 0.1$ and impurity-Bose coupling constant $\gib =  g n \xit$.
      }
    \label{fig:fig1}
\end{figure}

 While much theoretical work exists addressing the ground state properties of Bose polarons \cite{Rath2013,Levinsen-PRL2015,Casteels-PRA2014,Christensen-PRL2015,Grusdt2017b,Parisi2017a,Ichmoukhamedov2019,Panochko2019,Volosniev2017} there is still little understanding of finite temperature properties \cite{schmidt2016mesoscopic,levinsen2017finite,dzsotjan2020dynamical,Yan2019} and even more so of
 its non-equilibrium dynamics \cite{Catani2012,Meinert2017,boyanovsky2019dynamics,skou2021non,Koutentakis2022}. What happens when an impurity is injected into a weakly interacting Bose condensate?  What is the dynamics of the formation of a polaron and under what conditions and on what time scales can a stable quasiparticle form at all? We will address these questions in the present paper considering a point impurity interacting with a weakly interacting, one-dimensional Bose condensate in the full range of impurity-boson coupling strength, see \cref{fig:fig1}. The limit of weak impurity-boson interaction can be well described by the generation and subsequent binding or emission of Bogoliubov phonons from the impurity \cite{lausch2018} in terms of a Hamiltonian similar to that of the Froehlich model used in solid state systems \cite{Frohlich1954}. Most existing studies of the non-equilibrium dynamics of Bose polarons is based on this model \cite{Rath2013,Astrakharchik,Casteels2011,Shchadilova2016,Drescher2019,Ardila2021,Nielsen2019
}. It is however
 no longer well suited in the limit of strong impurity-boson coupling and thus we here follow a different approach. Starting from a full quantum description of the interaction of a mobile impurity with the condensate we employ a mean-field approach that takes the backaction of the impurity onto the condensate into account as in
 \cite{Hakim1997,Bruderer-EPL2008,Blinova2013}, but keeps the entanglement between impurity and BEC by working in a co-moving frame. This approach
 was shown to be very accurate for the prediction of ground state properties of Bose-polarons \cite{Jager2020} and bi-polarons \cite{Will2021} even for very strong impurity-boson couplings as long as the Bose-Bose interaction is weak. 
 Quantum effects are then taken into account by considering Bogoliubov excitations on top of the deformed condensate, which are here treated within a Truncated Wigner approximation \cite{Steel,Castin,C-field-review}. The advantage of this approach as compared to the 
 Froelich model and its extensions \cite{Shchadilova2016,Grusdt2017b,Ichmoukhamedov2019} is the substantially reduced number of deformed Bogoliubov phonons created by the impurity in such a description. As we will show the effect of these modified phonons can be neglected even in the non-equilibrium dynamics in many situations allowing for a comprehensive study of the polaron dynamics in terms of non-linear c-number differential equations.
 
\section{Energy-momentum relation 
in a homogeneous 1D Bose gas} \label{sec:model_and_stationary}

Before considering the time evolution of an impurity injected into a 1D Bose gas of neutral atoms, let us discuss the stationary properties of a polaron at finite momentum relative to the Bose gas.
A widely used approach to describe Bose-polarons for weak boson-boson and impurity-boson couplings is to consider the interaction of the impurity with Bogoliubov excitations of the unperturbed condensate. The resulting model is reminiscent of the Froehlich model in condensed matter physics. Due to the large compressibility of the Bose condensate, this model is however no
longer adequate if the impurity-boson coupling becomes large. In the latter case a growing number of phonons is generated at the location of  the impurity and boson-boson interactions become relevant. For this reason we here start from the full quantum model and apply a different approximation scheme.

\subsection{Model and modified mean-field approach}\label{sec:model}

A single  mobile impurity coupled to a Bose gas in a homogeneous one-dimensional system is described by the Hamiltonian
\begin{equation}
    \hH = \frac{\hp^2}{2M} +  \int \dd x\, \hphid{x} \Big[ -\frac{\partial_x^2}{2 m }  +  \half g \hphid{x} \hphi{x}  
    + \gib \delta(x- \hr) \Big] \hphi{x},\label{eq:H_homo}
\end{equation}
for $\hbar = 1$. Here $m$ ($M$) is the boson (impurity) mass, $\gib$  ($g$) the impurity-Bose (Bose-Bose) interaction constant, $\hphi{x}$ the bosonic field operator and  $\hr$ ($\hp$) the impurity position (momentum) operator. In the following we consider the case of repulsive coupling between all particles, i.e. $g,\gib>0$.

Since the system is homogeneous, its total momentum $\hp+\hPB$ is conserved, where $\hPB = -i \int dx \hphid{x} \partial_x \hphi{x}$ is the momentum of the Bose gas. The infinite homogeneous system is treated here as the limit of a finite system of length $L$ with periodic boundary conditions with $L\to \infty$.
In order to exploit the translational invariance  we apply the Lee-Low-Pines (LLP)  transformation \cite{Lee1953} $\hat{U}_\LLP = \exp( - i \hr \hPB)$, leading to the transformed Hamiltonian
\begin{equation}
\begin{aligned}
    \hH_\LLP &= \hat{U}_\LLP^\dagger \, \hH \, \hat{U}_\LLP \\
    & = \frac{1}{2M} \big(p - \hPB \big)^2  +  \int \dd x\, \hphid{x} \Big[-\frac{\partial_x^2}{2 m }  +  \half g \hphid{x} \hphi{x}  
    + \gib \delta(x) \Big] \hphi{x}. \label{eq:H_homo_LLP}
\end{aligned}
\end{equation}

Due to  translation invariance, $\hH_\LLP$ no longer depends on $\hat{r}$, and $\hp = \hat{U}_\LLP^\dagger (\hp+\hPB) \hat{U}_\LLP $ is the  conserved total momentum in this frame and can be replaced by a c-number $p$. We note that due to the LLP transformation, $\hat{\phi}^{(\dagger)}(x)$ describes the creation/annihilation of a boson in a frame co-moving with the impurity, such that $n(x) = \braket{\hphid{x} \hphi{x}}$ is the Bose gas density relative to the position of the impurity, or rather the relative impurity-boson density-density correlation function, to which we  refer from now on for simplicity as the Bose gas density.

Since $\hH_\LLP$ is an interacting many-body Hamiltonian, a complete solution of the dynamics is difficult without 
further approximations.
In \cite{Jager2020,Will2021} we have shown that for a weak boson-boson interaction, indicated by a small Tonks parameter $\gamma = g m / n $,  the ground state properties of a single or a pair of Bose polarons at rest is very well captured by a mean-field approximation that takes the backacktion of the impurity to the condensate into account and goes beyond the standard Froehlich model, which fails if $\gib \gg g n \xit $. Here $\xit = 1/\sqrt{2 g n \Tilde{m}}$ is the rescaled healing length with  the reduced mass $\tilde{m} = (1/m+1/M)^{-1}$. This motivates us to use this mean-field approximation 
also for the case of a moving impurity. We will test its validity by taking into account quantum fluctuations within a Truncated Wigner approach in \cref{sec:TWA}. 
The mean-field approximation amounts to 
replacing the field operator $\hphi{x}$ by a complex order parameter $\phi(x)$, whose
time evolution is determined by the non-linear Schrödinger equation \cite{Hakim1997}
\begin{equation}
    i \partial_t\; \phi(x,t) = \Big[ -\frac{\partial_x^2}{2 \tilde{m} }+ i\,  v(t)\,  \partial_x  + g |\phi(x,t)|^2 + \gib \delta(x) \Big] \phi(x,t) \label{eq:GPE_homo}.
\end{equation}
Here $v(t) = \big(p - \pB(t) \big)/M$ is the impurity velocity.

Since we are interested in the formation dynamics of polarons either after a sudden quench or an adiabatic turn-on of the impurity-Bose coupling constant $\gib$, the Bose gas is assumed to be initially in its ground state $\phi(x,t=0) = \sqrt{n}$ at $t=0$, where $n$ is the average density of bosons. As the Bose gas carries no initial momentum, the conserved total momentum is equal to the initial impurity momentum $ p  = M v(t=0)$. We simulate the time evolution with periodic boundary conditions numerically using a Fourier split-step method \cite{Geiser2019}. If not stated otherwise we choose the system size $L$ large enough such that signals are not able to reach the boundary for all times $t$ considered, i.e. $L \gg \tilde{c} t$, where $\tilde{c} = \sqrt{gn / \tilde{m}}$ is the rescaled speed of sound. 

\subsection{Stationary state and energy-momentum relation}\label{sec:stationary_state}

\begin{figure}%
    \centering
    \includegraphics[width=\textwidth]{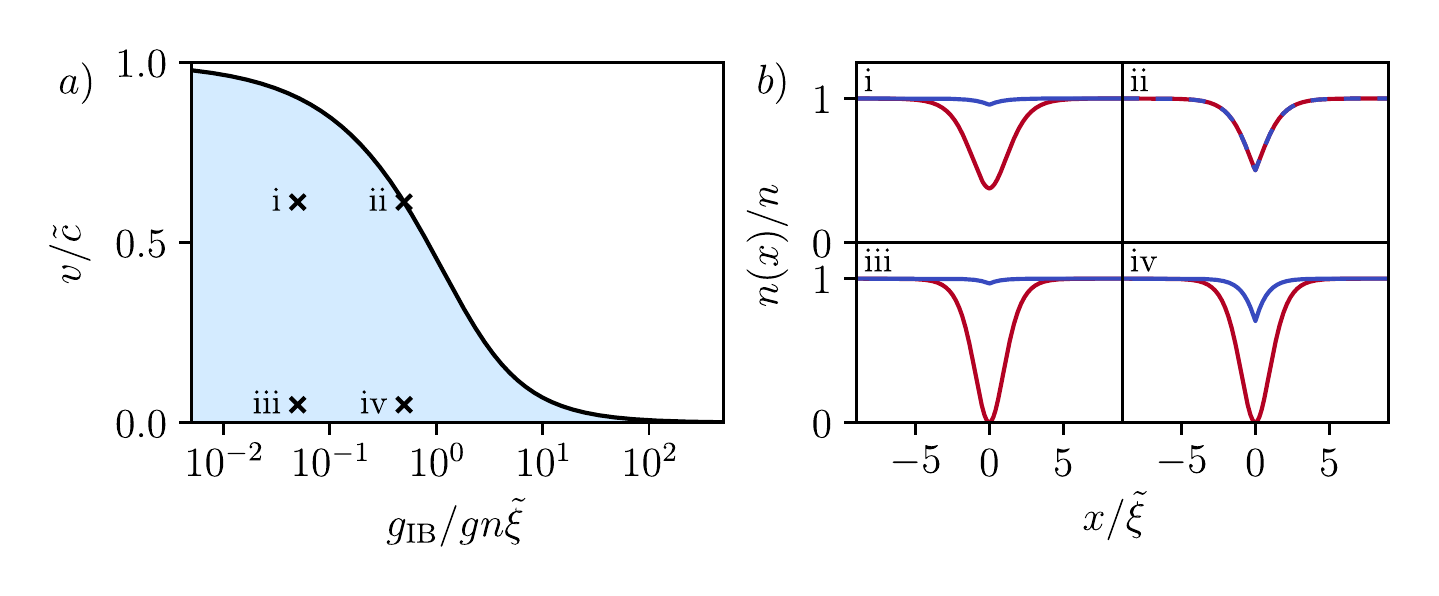}
    \caption{
       a) The black line shows the critical impurity velocity $v_c$. Only in the blue-shaded area, the system has two stationary states for each velocity $v$ and none otherwise. b) Density profile of the stationary states for different parameters, marked by crosses in a). At the critical velocity (ii) the two states are equal.
    }%
    \label{fig:state_ana}%
\end{figure}

We proceed by discussing the stationary properties of a polaron moving with non-zero total momentum $p$ by 
characterizing the steady-state solutions of \cref{eq:GPE_homo} \cite{Koutentakis2022}. Since the solution has a simpler form if the impurity velocity $v$ is used as a parameter, rather than the conserved momentum $p$, it is useful to derive the stationary state as a function of $v$ as in \cite{Hakim1997}, and then calculate the corresponding momentum by $p = Mv + \pB[\phi]$. This is possible since $v(t)=v$ is constant in the stationary state. 
The analytical expression of the state for a fixed $v$ was derived in \cite{Hakim1997,Jager2020,Tsuzuki1971,Volosniev2017}, see \ref{sec:app_stationary_state} for technical details and analytical expressions. 

As shown in \cite{Hakim1997},  below a critical impurity velocity $v_c$ the system has two stationary states for every given value of $v$, see \cref{fig:state_ana} b) for the corresponding density profiles. 
Above the critical value no stationary solution exists. Exactly at the critical velocity, the two states are equal two each other. 
$v_c$ depends on the coupling constant and can be found from the solution of \cite{Hakim1997,Volosniev2017}
\begin{equation}
    \frac{\gib}{g n \xit}  \overset{!}{=} \frac{1}{2 a_c} \sqrt{1-20a_c^2-8a_c^4 + (1+8a_c^2)^{3/2}}, \quad
    \text{where } a_c = v_c / \tilde{c}  \; . \label{eq:crit_velocity}
\end{equation}
  It is plotted in \cref{fig:state_ana} a). For small interaction $\gib \ll g n \xit$, it agrees with the prediction of the Froehlich model $v_c = \tilde{c}$ \cite{Landau1941,lausch2018}. However in the limit of strong interactions $\gib \gg g n \xit$ the condensate is strongly depleted at the impurity position, resulting in a vanishing critical velocity $v_c \to 0$.

\begin{figure}[ht]
    \centering
    \includegraphics[width=.99\textwidth]{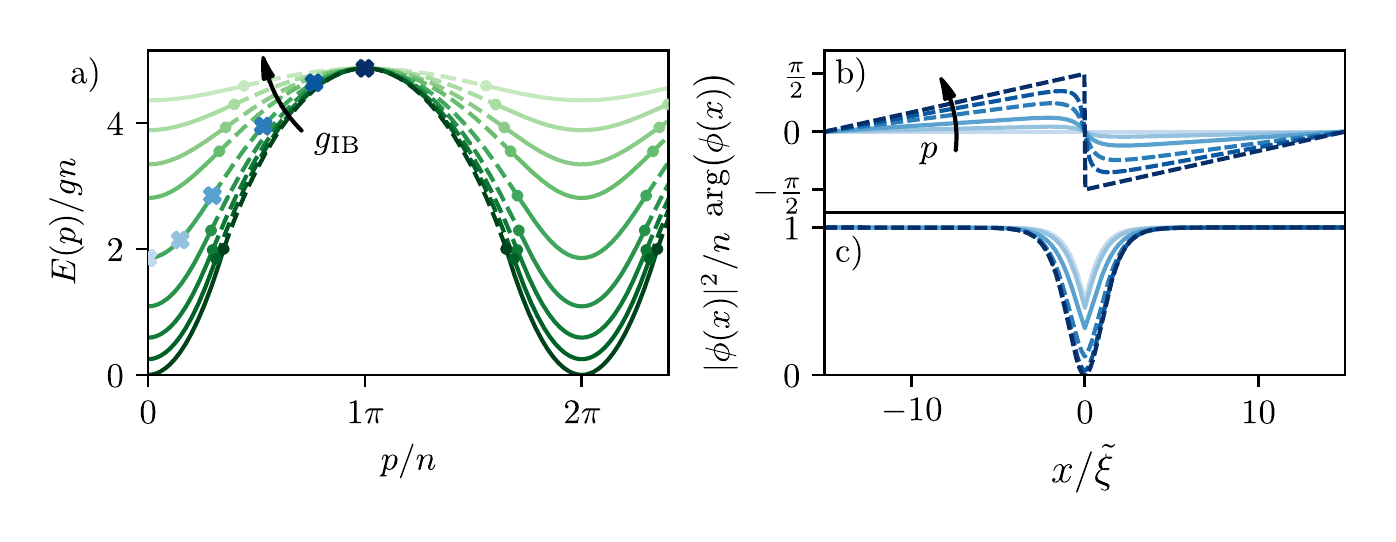}
    \caption{
    a) Energy momentum relation of the Bose polaron for different coupling constants $\gib/gn\xit =0, 0.1, \, 0.25, \,  0.5, \, 1, \, 2, \,  3, \, 5,\, 10$, the direction of the arrow points from weak to strong coupling. The mass ratio is $M = 3 m$ and Tonks parameter $\gamma=0.1$.  
    The phase/density of the Bose gas is shown in b)/c) for $\gib =  gn \xit$ and different total momenta $p$, below $p/n=\pi$ indicated by the blue crosses in a). Other parameters are as in a) except a finite system size $L=30 \xit$. For $\nu \pi \le p/n \le (\nu+1)\pi$ and integer $\nu$ the condensate picks up a constant phase gradient of $\nu\pi$ over its total length.} 
    \label{fig:dispersion}
\end{figure}

The polaron energy is given by the difference in energy of the full system with and without the impurity $E_\text{pol} = E(\gib,p) - E(\gib=0,p=0)$, where  $E(\gib,p)$ is the expectation value of the LLP Hamiltonian \cref{eq:H_homo_LLP} in a coherent state with amplitude $\phi(x,t)$.  $E_\text{pol}$  is shown as a function of the polaron momentum in \cref{fig:dispersion} a), where the two stationary states are distinguished by solid and dashed lines. It becomes clear that the polaron state is unique for every momentum $p$ and the two different states only refer to different parts of the energy-momentum relation. For the lower momentum state the energy-momentum relation was derived in \cite{Panochko2019}.

\begin{figure}[ht]
    \centering
    \includegraphics[width=.99\textwidth]{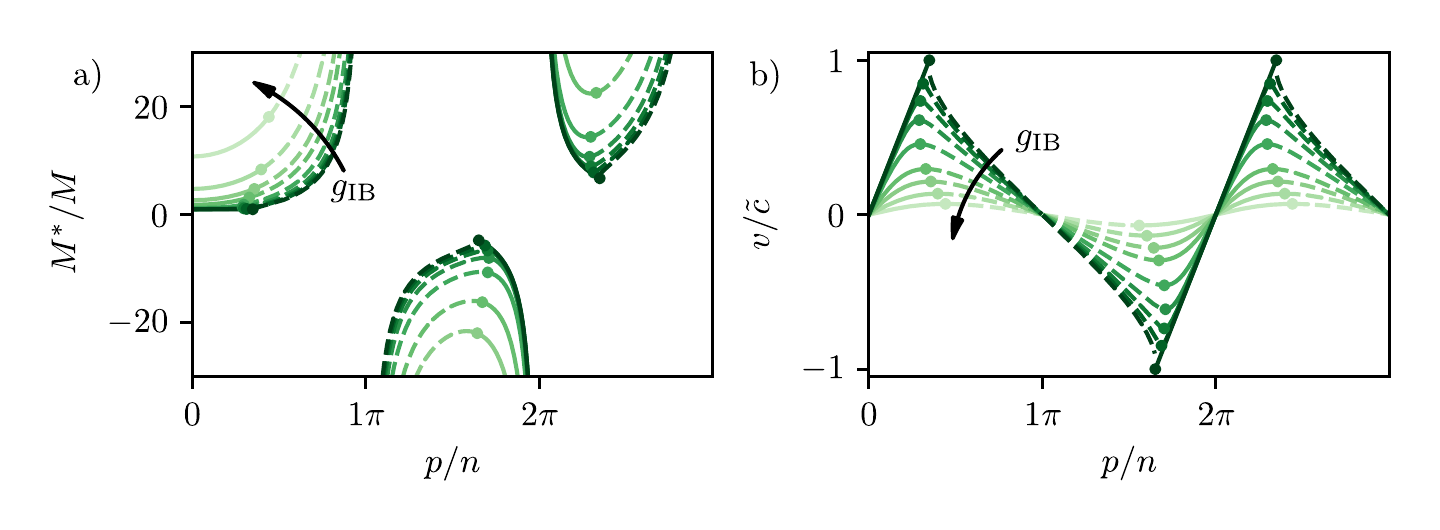}
    \caption{
    a) Kinetic polaron mass $M^*=p/v$ and b) velocity $v$  as a function of the momentum $p/n$ and for increasing coupling constants $\gib$ as in \cref{fig:dispersion} a) . The direction of the arrow indicates increasing $\gib$ in a) and b) for the different lines.
    }
    \label{fig:mass}
\end{figure}

From the existence of a critical velocity one might conclude that a Bose polaron exists only up to maximum polaron momentum $p_\text{max}$ depending on $g_\text{IB}$, and the energy momentum relation $E_\text{pol}(p)$ would terminates at some value of momentum. This is not the case. Instead when increasing $p$ further the solution smoothly crosses over into the second steady state with smaller kinetic velocity and larger condensate depletion. 
As can be seen from \cref{fig:dispersion} c) the condensate depletion grows with
increasing momentum. As a consequence the kinetic mass of the polaron, defined as the ratio of polaron momentum and velocity $M^*=p/v$, increases with $p$ as shown in \cref{fig:mass} a). When crossing from the momentum regime of the first solution of \cref{eq:GPE_homo}
to the second solution, the increase of the mass with momentum becomes larger than linear. This leads to a non-monotonous relation between polaron velocity $v$ and momentum $p$, plotted in 
 \cref{fig:mass} b). Note that the polaron velocity always stays below the weak-coupling critical value $v_c$. 

When the momentum reaches the value $p_\text{max}= n\pi $, the state of the condensate is exactly equal to a dark soliton, such that the energy is $E(p_\text{max}) = \tfrac{4}{3} n \tilde{c}$ and the impurity velocity goes to zero $v(p_\text{max})= 0$. In this case the condensate phase winds by $\pi$ over its entire length, which for periodic boundary conditions corresponds to 
half a flux quantum piercing through the ring. 
Since the density is fully depleted the kinetic polaron mass $M^*(p_\text{max})$ diverges at this point. 

We note that the relation between momentum, energy, and velocity of the polaron has already been found for momenta $|p|< n \pi$ \cite{Koutentakis2022}. A peculiar behavior of $E(p)$ is however seen when the total momentum is increased further. The energy starts to decrease with increasing momentum, corresponding to a negative group velocity $\partial E/\partial p$ and consequently a negative
impurity velocity. In fact, as can be seen from \cref{fig:dispersion}a, $E(p)$ is a periodic function of $p/n$ with period $2\pi$. This is because the properties of the polaron 
are determined by collective excitations of the Bose gas, whose energy-momentum relation in 1D is periodic with period $2\pi n$ due to Luttingers theorem. Here it can be interpreted as follows:
If the total momentum is in the range $(\nu-1)\pi \le  p/n\le \nu \pi$, with $\nu$ being an integer, the condensate picks up an integer winding of its phase over its whole length (period) in the stationary state.
If $\nu$ is even, the momentum picked up by the background condensate exceeds the total momentum and the excess must be compensated by a relative motion of the impurity against the condensate, corresponding to negative values of $v$. 

We will see in the following that the periodic behavior of the energy-momentum relation can give rise to negative asymptotic impurity velocities after injecting it into the condensate with large positive initial velocity. We note that the reversal of the impurity velocity has already been seen in experiments with a strongly interacting Bose gas ($\gamma \gg1$), where Bloch oscillations of an impurity subject to a constant force have been observed \cite{Meinert2017}.



\section{Mean-field description of polaron formation}\label{sec:homogeneous_dynamic}

We now discuss the dynamics when an impurity is injected into a homogeneous condensate with finite initial momentum. First we consider a quasi-adiabatic turn-on of the impurity-boson coupling and subsequently discuss a sudden quench.
\subsection{Quasi-adiabatic evolution}\label{sec:adiabatic}
Let us first investigate the dynamical properties of the system when the impurity-boson coupling constant $\gib(t)$ is turned on slowly compared to the other time scales of the system. We will show that even though the energy spectrum of the full system is gapless in the thermodynamic limit, a local adiabatic following of the polaron ground state is possible if the initial velocity of the impurity is subsonic.

\begin{figure}[ht]
    \centering
    \includegraphics[width=.99\textwidth]{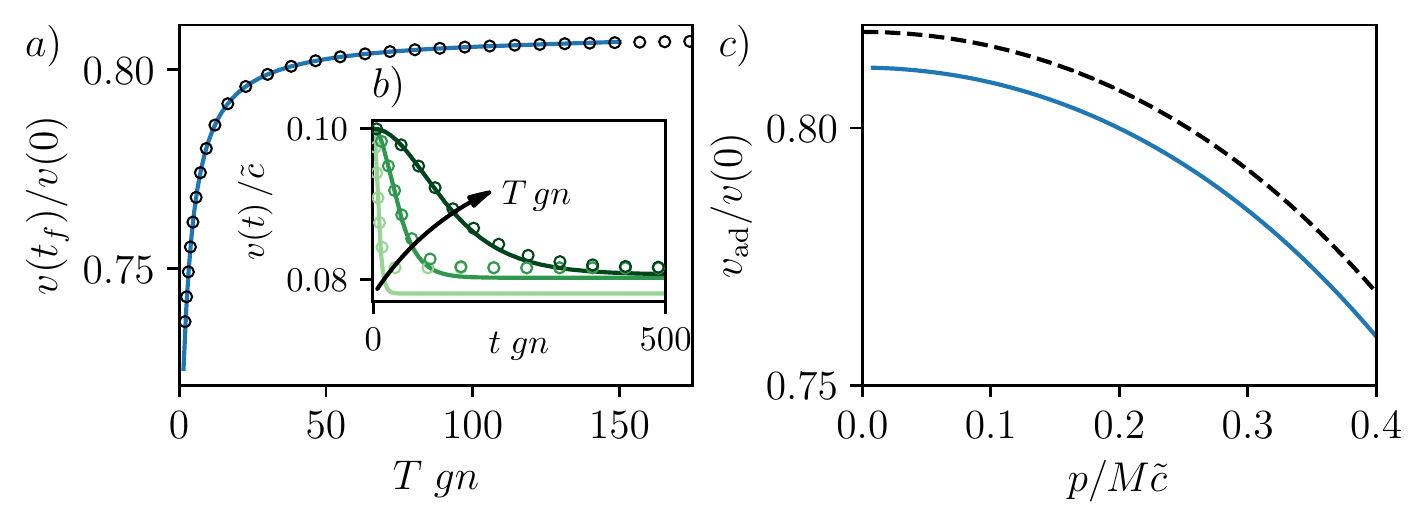}
    \caption{
    a) Final impurity velocity $v(t_f)$ after the turn-on with timescale $T$ of the impurity-Bose coupling constant. The initial velocity is $v(0) = p/M = 0.1 \tilde{c}$ and we choose a mass ratio of $M=3m$, Tonks parameter $\gamma =0.1$ and $\gib = gn \xit$. The numerical result (blue line) agrees very well with the exponential fit (black circle) of $\alpha_1 e^{-\alpha_2/T} + \alpha_3$, where $\alpha_i$ are fitting parameters. The simulation is performed up to $t_f = 600 /gn$. 
    b) Time evolution of the impurity velocity for $ T gn = 10, 50, 150$, indicated by the black arrow. Lines are simulated results and circles are analytically calculated from the instantaneous stationary state. 
    c) Final impurity velocity $v_\text{ad}$ in the adiabatic limit $T \gg 1/gn$ for different total momenta $p = M v(0)$. Simulated result (blue solid) differ only slightly from the analytical prediction (black dashed). The deviations are caused by the system not being gaped, see \ref{sec:app_adiabatic2}.}
    \label{fig:adiabatic}
\end{figure}

\begin{figure}[ht]
    \centering
    \includegraphics[width=0.9\textwidth]{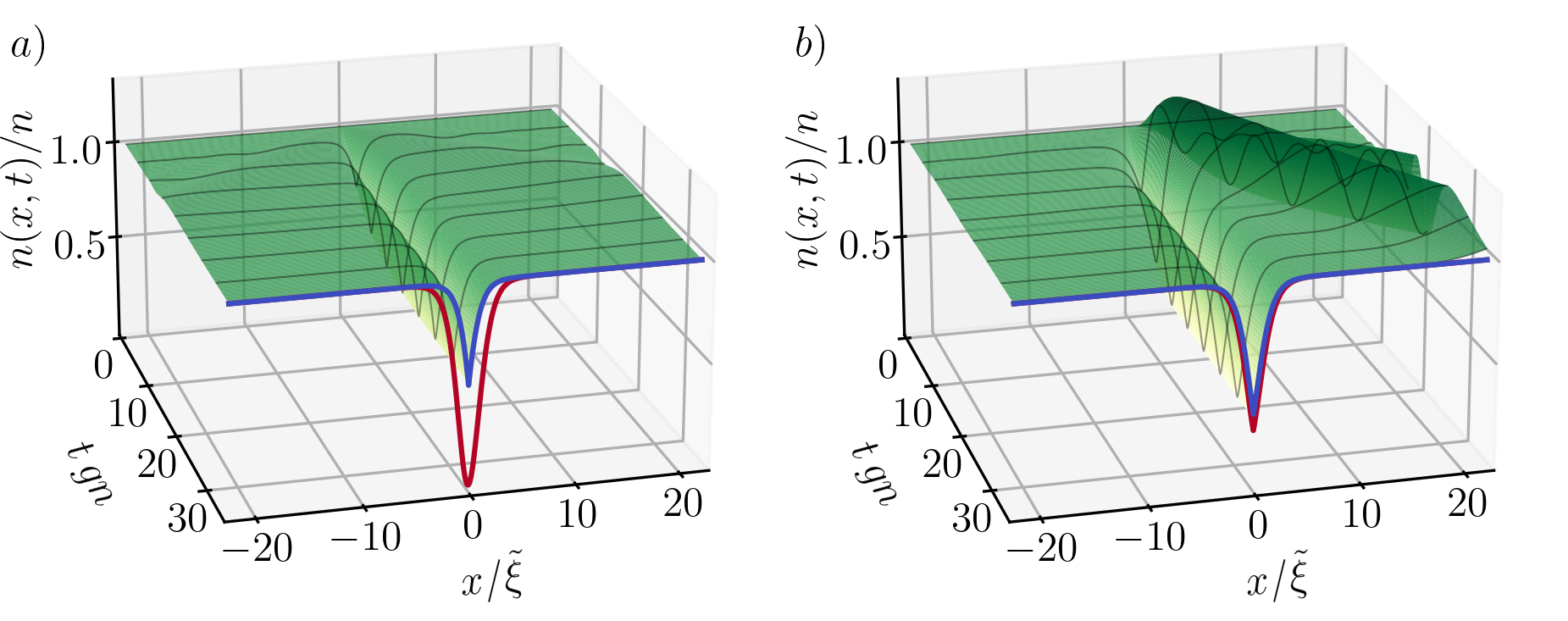}
    \caption{
        Evolution of the Bose gas density close to the impurity  for an a) initially slow $v(0) = 0.1\, \tilde{c}$ and b) fast $v(0) = 1.1\, \tilde{c}$ impurity. The impurity Bose coupling constant of strength $\gib = gn \xit$ is turned with the timescale $T = 10 /gn$. The mass ratio is constant at $M=3m$ and the Tonks parameter is $\gamma = 0.1$.
        The blue and red lines are the two analytically derived stationary states evaluated at the final impurity velocity $v(t)$. The system converges locally in either one of the two states, dissipating the energy as density waves.
    }
    \label{fig:density_example}
\end{figure}

In order to achieve a smooth turn-on protocol, we choose a time dependence of the coupling according to:
\begin{equation} \label{eq:gib_t}
    \gib(t) = \gib\, \tanh(t/T). 
\end{equation}
$\gib$ is the final coupling constant and $T$ the turn-on timescale, which we chose large compared to the inverse chemical potential $1/gn$. Since the critical velocity $v_c$, below which stationary states exist, depends on $\gib$ it is also time-dependent and the time evolution differs qualitatively whether the impurity momentum is below or above $v_c$ at any time. The time evolution of the condensate for the two cases is shown exemplarly in \cref{fig:density_example}, where the blue and red lines indicate the two stationary states corresponding to the final velocity of the impurity.

First, we focus on the case of a slow impurity $v(t) < v_c(t)$ for all $t$, such that \cref{eq:GPE_homo} has a stationary solution for all $v(t)$. The evolution of the impurity velocity is exemplary shown for different $T$ in \cref{fig:adiabatic}b). Even though the impurity is always subsonic it is decelerated to a finite value $v(t_f)<v(0)$, which increases monotonously with $T$, see \cref{fig:adiabatic}a). Notably also in the limit $T \to \infty$ the impurity is still slowed down and $v(t_f)$ does not converge to $v(0)$. 
This  is caused by the formation of the polaron. In the instantaneous ground state, the total conserved momentum $p$ is the polaron momentum and is related to the impurity velocity by the effective mass $m^*$ off the polaron 
\begin{equation}
    v(t) = \frac{p}{m^*(t)} .\label{eq:v}
\end{equation}
Since the effective mass increases monotonously with the coupling constant $\gib$ \cite{Jager2020}, the impurity must decelerate when the impurity-Bose coupling constant is turned on. For finite turn-on times $T$,  density waves are created during the formation of the polaron leading to an additional friction force.

To quantify the quasi-adiabatic slow-down we fit an exponential $\alpha_1 e^{-\alpha_2/T} + \alpha_3$ to the simulated impurity velocity (see \cref{fig:adiabatic}a). From the fit the final impurity velocity in the adiabatic limit is determined by $v_\text{ad} = \alpha_1 + \alpha_3$. It is shown as a function of the conserved total momentum $p = M v(0)$ in \cref{fig:adiabatic}c), where it becomes apparent that the deceleration occurs for all $p$. Due to the gaplessness of the system to collective excitations the simulated final velocity is always slightly below the quasi-adiabatic value, following from \cref{eq:v} with $m^*$ replaced by the effective mass of the stationary solution, see \cref{fig:mass}. This is because the adiabatic theorem \cite{Nenciu1980} strictly does not hold. Assuming strict adiabatic following we can derive the relation between $v(t)$ and $p$ as a function of the instantaneous coupling constant $\gib(t)$ in the polaron ground state (see \ref{sec:app_stationary_state}). \cref{fig:adiabatic}a) shows that the simulated time evolution follows the instantaneous ground state reasonably well for large $T$ except for the small difference shown in\cref{fig:adiabatic}c), see \ref{sec:app_adiabatic2} for details. 

\begin{figure}[ht]
    \centering
    \includegraphics[width=0.99\textwidth]{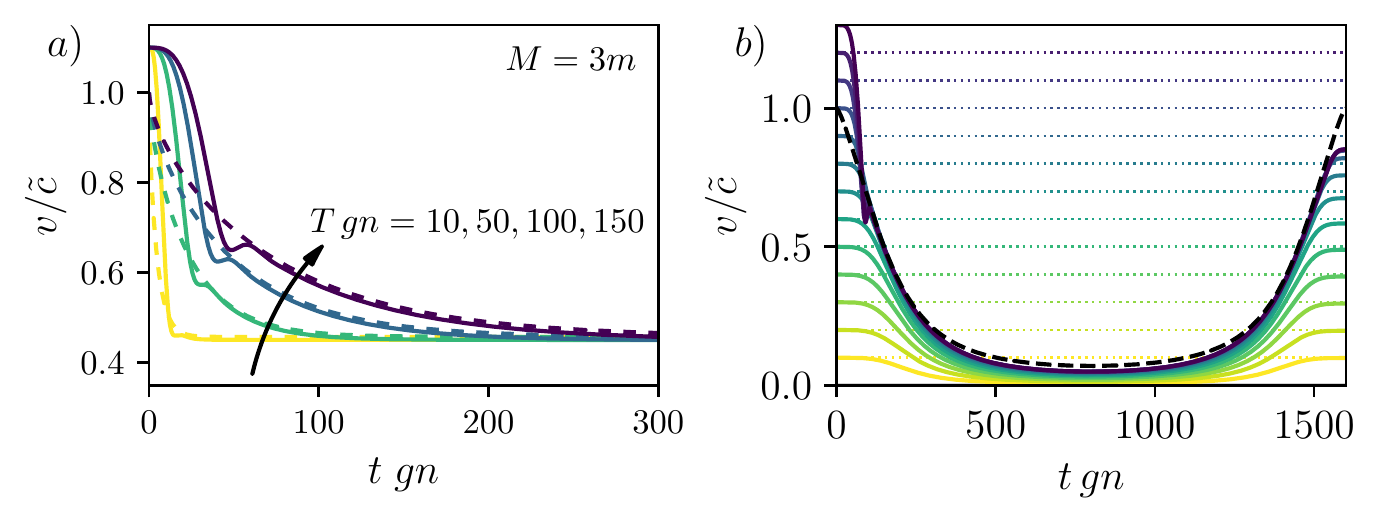}
    \caption{
    a) Time evolution of the impurity velocity (solid line) for a supersonic impurity $v(0) = 1.1 \tilde{c}$ and different turn-on times $T$. The initial fast deceleration slows down as soon the impurity is below the critical velocity $v_c( \gib(t))$ (dashed lines). The mass ratio is $M=3m$, the Tonks parameter $\gamma=0.1$, and the final impurity coupling is $\gib = gn \xit$.
    b) Evolution of the impurity velocity for different initial velocities. The coupling constant is switched on and off again by  $\gib(t) = \gib \sin^2(\half \pi \,t /T)$, with $\gib = 10 gn \xit$, $T = 800$ and otherwise parameters as in $a)$. The black dashed line is the instantaneous critical velocity and the dotted lines mark the initial velocity. The deceleration of subsonic impurities is reversible, but not for supersonic ones.}
    \label{fig:non_adiabatic}
\end{figure}

We now consider an impurity that is initially faster than the critical momentum $v(0) > \tilde{c}$. In that case, \cref{eq:GPE_homo} does not have a stationary solution in the initial phase, which leads to the creation of density waves \cref{fig:density_example}b) and a friction force acting on the impurity. \cref{fig:non_adiabatic}a) shows that the impurity is  quickly decelerated until its velocity is below the critical $v_c(\gib)$. Afterward, the system again  follows the instantaneous ground state quasi-adiabatically, resulting in a slower deceleration. 

An important difference between these two processes is, that the second slow deceleration, related to the adiabatic formation of the polaron, is reversible, while the other one is not. This can be seen in \cref{fig:non_adiabatic}b), where the impurity-Bose coupling constant is turned on and off again by $\gib(t) = \gib \sin^2(\half \pi \,t /T)$. Here impurities that are initially below the critical momentum are almost brought to a standstill when the coupling constant is at a maximum but accelerate again to the initial velocity when $\gib$ is turned off again. In contrast, impurities starting above criticality are not reaching their initial momentum again after the sweep. The small deviation in the final velocity for sub-critical trajectories is again caused by the system not being gaped in the thermodynamic limit.

\subsection{Quench}

Here we examine the evolution of the system when the coupling constant is abruptly quenched at $t=0$.

\begin{figure}[ht]
    \centering
    \includegraphics[width=\textwidth]{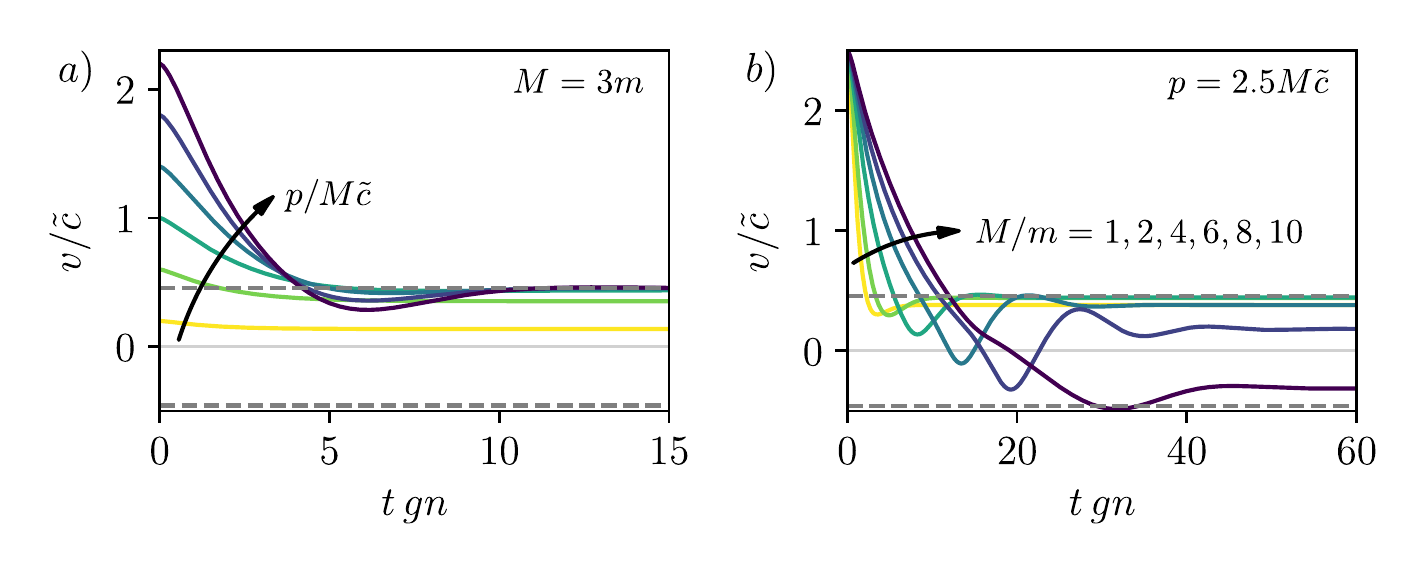}
    \caption{
    Polaron quench dynamic for a) different initial impurity velocities and b) different mass ratios. We choose a coupling constant of $\gib = gn \xit$ and $\gamma = 0.1$. The horizontal dashed grey lines are $\pm v_c$.  The impurity velocity always converges to a value $|v(t) | < v_c$. 
    }
    \label{fig:velocity_quench}
\end{figure}

The time evolution of the impurity velocity after a sudden turn-on of the interaction with the condensate is shown in \cref{fig:velocity_quench}a) and b) for different initial velocities and
different ratios of bare impurity mass $M$ to boson mass $m$.
The quench leads to radiation of density waves until the system reaches a steady state. For all initial conditions and parameters, a friction force \cite{Koutentakis2022} is exerted on the impurity and slows it down until the final velocity is reached smaller than the critical $v_c$. This agrees again with the analytic prediction, that the system has a stationary state only below $v_c$. However, for a large initial momentum, the impurity velocity is non-monotonic, which cannot be explained by a frictional force alone.

If the impurity is heavy a rather unexpected behaviour is found for sufficiently large initial momentum, see  \cref{fig:velocity_quench} b).  First, as expected the deceleration is slower in the case of heavier impurities. However for a sufficiently 
large mass, the impurity is not only slowed down, but the velocity can change its direction before converging to a constant velocity. In this case, the background condensate attains an additional momentum by building up a finite phase gradient away from the impurity locally approaching a stationary polaron solution with negative impurity velocity as discussed in the previous section.

\begin{figure}
\centering
\includegraphics[width=\textwidth]{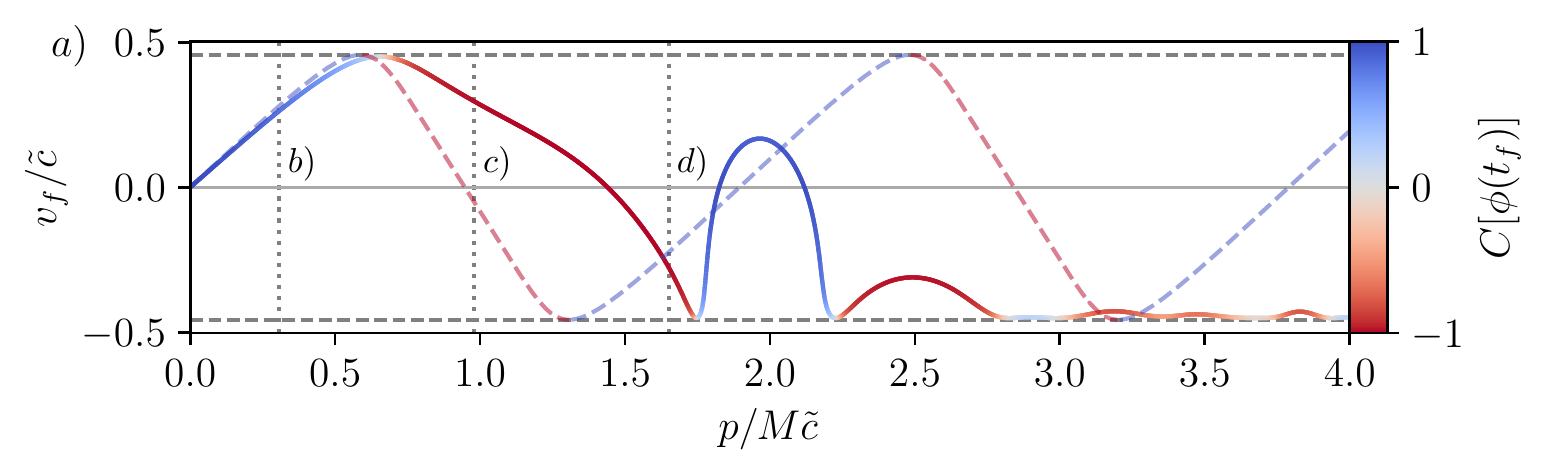}
\caption{
    a) Final velocity $v_f$ after a quench of the impurity-Bose coupling constant to $\gib = gn \xit$ for different total momenta $p=M v(0)$. For the mass ratio we choose $M = 10 m$ and the Tonks parameter $\gamma =.1$.  The final velocity is always in an interval $[-v_c,v_c]$, marked by grey vertical lines. The color code indicates the generalized contrast \cref{eq:contrast}. The evolution of the Bose gas densities at the marked positions is shown in \cref{fig:fig1} b)-d). The colored dashed line shows the velocity-momentum relation of the stationary state. It agrees well with the simulated result for sub-critical initial velocities $v(0) < v_c$.  For higher total momenta emission of density waves or solitons leads to initial friction forces, causing the deviation.
} \label{fig:state_population} 
\end{figure}

This effect is examined in more detail in \cref{fig:state_population} a). It shows the final impurity velocity as a function of the conserved momentum $p = M v(0)$ of a heavy impurity $M = 10 m$. Again the final velocity is for all initial conditions in the interval $[-v_c,v_c]$. However, as mentioned in \cref{sec:stationary_state} the system has two stationary states for each velocity and we examine which of these states are populated. For this we first focus on \cref{fig:fig1} b) - d), showing the evolution of the Bose gas density for different total momenta. The red and blue lines are the analytically calculated stationary states, where we used the final impurity velocity from the simulation as a parameter, rather than the total momentum. Depending on the initial conditions the system converges in either of these two states. In order to quantify the overlap of the final polaron state with either of the two stationary states we determine the generalized contrast
\begin{equation}
\begin{aligned}
    C[\phi(t)] &= \frac{|\braket{\phi(t)|\phi_1} | - |\braket{\phi(t)|\phi_2} | }{|\braket{\phi(t)|\phi_1} | + |\braket{\phi(t)|\phi_2} | -2 |\braket{\phi_1|\phi_2} | } \\
    \braket{\phi_a|\phi_b} &= \int_{-l/2}^{l/2} \dd x\; \phi_a(x)^* \phi_b(x). \label{eq:contrast}
\end{aligned}
\end{equation}
Here $\ket{\phi(t)}$ is the simulated state of the system and $\ket{\phi_i}$ the steady states, evaluated at the simulated final velocity. The contrast is $1$ if the system converges into the first state and $-1$ for the second. This definition differs from the standard expression of contrast by the term $\braket{\phi_1|\phi_2} $ in the denominator, which we need since the stationary states are not orthogonal. We evaluate the scalar products over an interval $l$ which we choose such that the time evolution is properly converged within it. 
The contrast is depicted in \cref{fig:state_population} a) by the color code and it becomes apparent that the polaron changes its state when its velocity intersects with the critical value $v_c$. This is explained by the states being equal at the critical velocity.

The explanation for the non-monotionic impurity velocity and the change in the movement direction at larger total momenta can be seen in \cref{fig:fig1} d). At large momenta, the impurity carries enough energy that a grey soliton is created besides density waves which carry additional momentum. This is visible as the local depletion moving away from the impurity at a slower speed as the initially created density waves in \cref{fig:fig1} d). Note, however, that this interpretation is based on numerical evidence only where a non-trivial behaviour of the impurity velocity always coincided with soliton emission.

\section{Truncated-Wigner approximation of a harmonically trapped polaron}
\label{sec:TWA}

The above mean-field analysis has neglected quantum fluctuations. Although it has been shown in \cite{Jager2020,Will2021} that the ground state
properties of Bose polarons and bi-polarons are well described by the modified mean-field approach even in the limit of strong boson-impurity 
coupling $\gib \gg g n \xit$, provided the Bose-Bose interaction is weak, i.e. if $\gamma\ll 1$, it is not clear if this still holds in the non-equilibrium case. For this reason we now consider the effect of small quantum fluctuations using a truncated Wigner approach 
\cite{Steel,Castin,C-field-review}. This approach fully captures the influence of (deformed) Bogoliubov phonons on top of the condensate in quadratic (Bogoliubov) approximation. However, as stated by the Mermin-Wagner-Hohenberg theorem \cite{mermin1966absence,hohenberg1967existence} there is no true Bose condensation in homogeneous 1D gases, which manifests itself by infrared divergencies when considering lowest-order quantum fluctuations. The latter also holds for finite systems with periodic boundary conditions. Thus in order to describe  quantum fluctuations we can no longer approximate the one-dimensional gas as being homogeneous and have to take into account the presence of a harmonic trapping potential. 
(The different regimes of quantum degeneracy in trapped 1D Bose gases are discussed e.g. in \cite{Petrov2000}.)
This complicates the theoretical description as the Lee-Low-Pines transformation, conveniently used in homogeneous systems, no longer leads to a decoupling of the total momentum of the polaron. We will show however that the total momentum obeys a simple equation of motion if also the impurity is trapped.

\subsection{Lee-Low-Pines Hamiltonian of a trapped 1D Bose gas}

We start by adding a harmonic potential with frequency $\omega$ for bosons and $\Omega$ for the impurity to the Hamiltonian \cref{eq:H_homo} to avoid infrared divergencies in the Bogoliubov theory of boson-boson interactions. 
Since these potentials break the translation invariance of the system the total momentum of the system is no longer conserved. 
Therefore, the transformation into a relative and a center of mass coordinate by the LLP transformation does not eliminate the impurity operators.
Nevertheless, it is useful to apply the transformation, since $\hp$ and $\hr$ only appear up to quadratic order in the LLP Hamiltonian
\begin{equation}
\begin{aligned}
    \hH_\LLP &= \frac{1}{2M} \big(\hp - \hPB \big)^2 + \half M \Omega^2 \hr^2 \\
    & +\int \dd x\, \hphid{x} \Big[ -\frac{\partial_x^2}{2 m }   + \half m \omega^2 (x+ \hr )^2 +    \half g \, \hphid{x} \hphi{x}  + \gib \delta(x) \Big] \hphi{x}.
\end{aligned}
\end{equation}
To simulate the time evolution of the system we derive the Heisenberg equation of motion $\partial_t\, \bullet = i\; [\hH_\LLP \, , \, \bullet \,]$ for $\hp(t)$, $\hr(t)$ and $\hphi{x,t}$. The advantage of the LLP transformation even in the case of harmonic trapping becomes clear here since the equations for $\hp(t)$ and $\hr(t)$ are formally solvable and we get
\begin{align}
\hp(t) &= \hp(0) \cos{ \Omega t} + \frac{1}{\Omega} \dot{\hp}(0) \sin{ \Omega t}
    + \frac{\Omega^2 - \omega^2}{\Omega}  \, \int_0^t \dd t^\prime \sin \Big( \Omega (t-t^\prime) \Big) \hPB(t^\prime), \label{eq:trap_p_sol}\\
    \hr(t) &= - \frac{\dot{\hp}(t) + N m \omega^2 \,\hXB(t) }{ M \Omega^2 + N m \omega^2}. \label{eq:trap_r_sol}
\end{align}
Here $N$ is the particle number of bosons and $\hXB(t) = \int \, \dd x\,  x \, \hphid{x,t} \,  \hphi{x,t} / N $ their center of mass position. In the case of equal trapping frequencies $\omega = \Omega$, the last term in the solution of the total momentum $\hp(t)$, \cref{eq:trap_p_sol}, vanishes, and its time evolution corresponds to that of an uncoupled harmonic oscillator
\begin{equation}
    \hp(t) = \hp(0) \cos{ \Omega t} + \frac{1}{\Omega} \dot{\hp}(0) \sin{ \Omega t}, \qquad \text{for} \quad \Omega= \omega. \label{eq:trap_p_free}
\end{equation}
The remaining equation for the bosonic field $\hphi{x,t}$ then reads:
\begin{equation}
\begin{aligned}
    i \partial_t \hphi{x,t} = \biggl\{  &-\frac{\partial_x^2}{2 m } + \frac{i}{M} \big[ \hp(t) - \hPB(t) \big] \partial_x \\
    &+ g \, \hphid{x,t} \hphi{x,t} + \gib \delta(x) 
    +\half m \omega^2 \big[x+ \hr(t) \big]^2 \biggr\} \hphi{x,t}. \label{eq:heisbenberg_phi}
\end{aligned}
\end{equation}
%

\subsection{Truncated Wigner simulation of the boson field}

We now solve the system of \cref{eq:trap_p_sol,eq:trap_r_sol,eq:heisbenberg_phi} in a limit where quantum fluctuation of the impurity position and momentum  are negligible, but include fluctuation of the Bose field using a truncated Wigner phase space approach (TWA). For reviews on the TWA methods see \cite{Castin,C-field-review}.  
To be able to apply a TWA we have to first treat the quantum evolution of $\hr(t)$ and $\hp(t)$.
For this we apply another approximation and replace the total momentum and position operator of the Bose field in the dynamical equations of the impurity by expectation values 
\begin{equation}
    \hPB \rightarrow \braket{\hPB} \coloneqq P_\text{B},\quad\textrm{and}\quad \hXB \rightarrow \braket{\hXB} \coloneqq X_\text{B}.
\end{equation}
This approximation is reasonable since the number of bosons is large $N \gg 1$ such that fluctuations of their center of mass coordinate are small. From \cref{eq:trap_p_sol,eq:trap_r_sol} then follows, that the fluctuation of $\hp$ and $\hr$ do not grow in time. It is possible to prepare the impurity in a state where fluctuation are small as long as its harmonic oscillator length scale $l_I = 1/ \sqrt{\Omega M}$ is small when compared to all other length scales of the system, especially the healing length of the condensate $\xi = 1/ \sqrt{2 gn m}$, where $n$ is the peak density of the ground state of the trapped Bose gas without an impurity. The semiclassical treatment of impurity position and total momentum is therefore justified only in the regime where $\frac{gn m}{\Omega M} \ll 1$. Under this condition the operators 
\begin{equation}
    \hr \rightarrow \braket{r} \coloneqq r,\quad\textrm{and}\quad \hp \rightarrow \braket{p} \coloneqq p
\end{equation}
are replaceable by expectation values.

In order to calculate the time evolution of the Bose field in a Wigner phase space description, an expression for the initial ground state of the Bose gas including quantum fluctuation is needed. Since we consider a weakly interacting Bose gas, it suffices to do this by replacing the field operators $\hphi{x,t=0}$ by the mean-field ground state $\phi_0(x)$ of a trapped Bose gas and add quantum fluctuation within Bogoliubov-de Gennes (BdG) approximation
\begin{equation}
\begin{aligned}
    \hphi{x,t} = \phi_0(x+r) + \sum_n \Big[u_n(x+ r)  \hat{b}_n(t) + v_n(x +r)^* \hat{b}^\dagger_n(t) \Big].
\end{aligned}
\end{equation}
Here $u_n(x)$ and $v_n(x)$ are the BdG coefficients of the trapped Bose gas and $\hat{b}^{(\dagger)}_n$ the phonon operators of the respective modes, see \ref{sec:BdG} for more details. The position $r$ appears in this expression since the Bose gas ground state is transformed into the LLP frame, corresponding to the shift by $r$. In the Wigner phase-space description, the phonon operators are replaced by stochastic c-numbers $\hat{b}^{(\dagger)}_n(t) \rightarrow \beta^{(*)}_n(t)$ where all stochasticity is in the initial state. Since this state is the phonon vacuum they are set to Gaussian random variables with mean and variance given by $\braket{\beta_n(0)} = 0 $ and $\braket{\beta_n (0)\beta_m^*(0)} = \half \delta_{n,m} $, corresponding to a virtual occupation of half a phonon per mode on average. By symmetric ordering of the Heisenberg equation \cref{eq:heisbenberg_phi} and replacing the operators by c-numbers $\hat\phi \to \phi$ we get the c-number equation of motion
\begin{equation}
\begin{aligned}
    i \partial_t \phi(x,t) = \Big\{  &-\frac{\partial_x^2}{2 m } + \frac{i}{M} \big[ p(t) - P_\text{B}(t) \big] \partial_x \\
    &+ g \, |\phi(x,t)|^2  - 2 g \, n^w\big(x+r(t)\big)  + \gib \delta(x)\\
    &+\half m \omega^2 \big[x+ r(t) \big]^2 \Big\} \phi(x,t). \label{eq:TWA_equation}
\end{aligned}
\end{equation}
Here $n^w(x)$ is the virtually occupied particle density due to the Wigner description and is given by
\begin{equation}
     n^w(x) = \half \sum_n^{n_\text{max}} |u_n(x)|^2 - |v_n(x)|^2.
\end{equation}
Note that we have truncated the number of modes taken into account, which is necessary in TWA, since if all BdG modes are included $n^w(x)$ would diverge $n^w(x) = \half \delta(0)$. The truncation of higher modes is commonly used in TWA simulations of trapped gases \cite{Castin,Steel} and is physically justified as quantum fluctuations of high frequency modes can be neglected. We simulate \cref{eq:TWA_equation} multiple times for different initial conditions and average about the different realization. In order to obtain expectation values all operators need to be symmetrically ordered first, e.g. the Bose gas density is given by
\begin{equation}
    n(x,t) =  \Bigl \langle  \big|\phi(x-r(t)\, ,t)\big|^2  - n^w(x) \Bigr \rangle,
\end{equation}
where $r(t)$ appears, such that the expression describes the density in the laboratory and not LLP frame.

\begin{figure*}[ht]
    \centering
    \includegraphics[width=\textwidth]{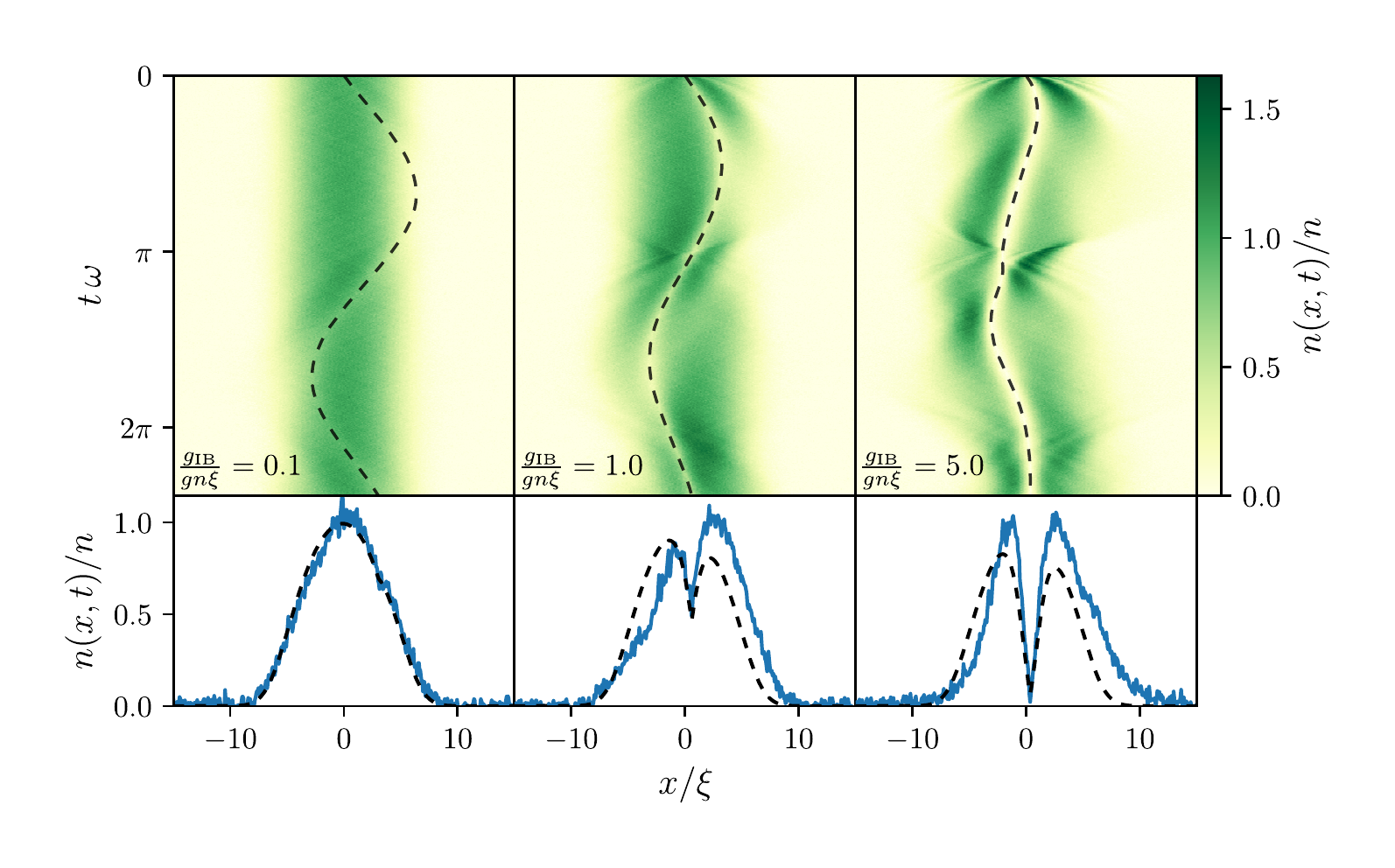}
    \caption{
    TWA simulation of a trapped system for different impurity-Bose coupling constants. Here trapping frequencies for bosons and impurity are equal $\Omega= \omega = 0.3 gn$, the Tonks parameter is $\gamma = 0.1$, the mass ratio is $M/m = 10 $ and the initial impurity momentum is $p = 0.75 M c$. The upper panels show the time evolution up to $t = 25/gn = 7.5/\omega$, where the dashed line marks the position of the impurity. The lower panels show the final density (blue) and compare it to the stationary state of the homogeneous system (dashed) modulated by the initial mean-field density of the trapped gas. The TWA simulation is averaged over 5000 noise realizations. 
    }
    \label{fig:trap_density}
\end{figure*}

The time evolution of the trapped system is illustrated in \cref{fig:trap_density}. A problem of this approach is that the system is finite, so that density waves created by the impurity oscillate in the trap and return back to the impurity. The system does therefore only reach locally a stationary state in very shallow traps
with very small trapping frequency $\omega \ll gn$. This however conflicts with the condition of a classical impurity $\frac{gn m}{\Omega M} \ll 1$ for reasonably heavy impurities and equal trapping potential.  
However, although the system is not in a stationary state the stationary solution described in \cref{sec:stationary_state} agrees reasonably well with the observed density distribution, see \cref{fig:trap_density} when applying a local density approximation. 

\begin{figure*}[ht]
    \centering
    \includegraphics[width=\textwidth]{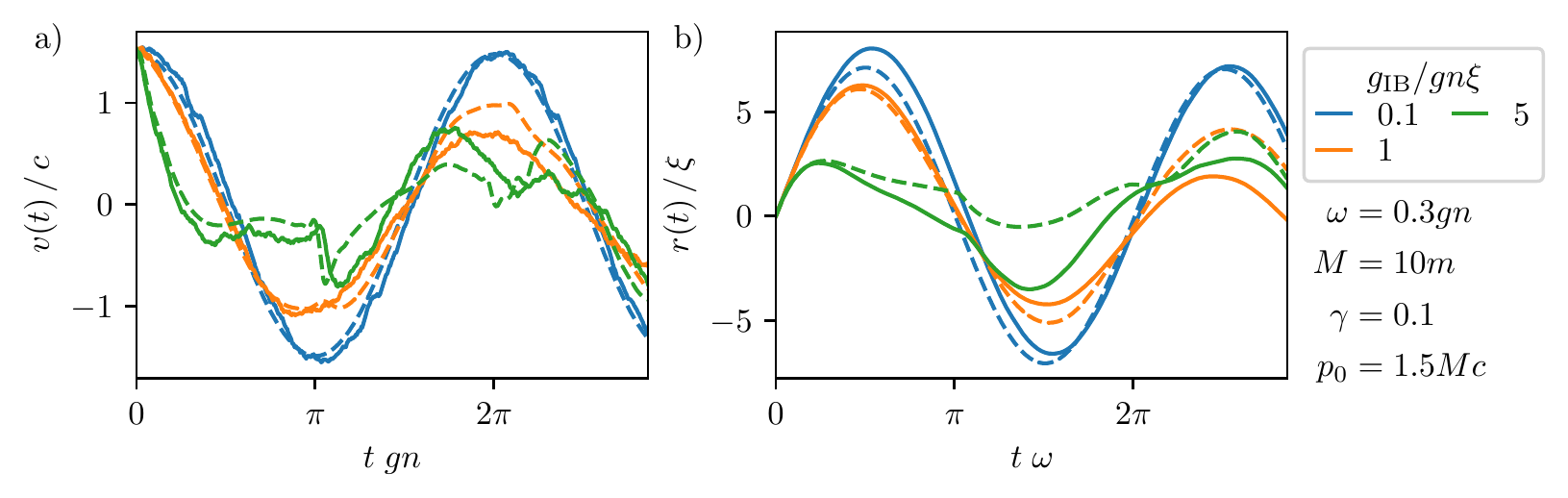}
    \caption{a) Impurity velocity and b) impurity position in a trapped system with equal trapping frequencies $\Omega = \omega$. Solid lines show the TWA simulation,  dashed lines the classical mean-field approximation. The TWA simulation is averaged over 5000 noise realizations.   } 
    
    \label{fig:trap_p_r}
\end{figure*}

Next, in order to test the validity of the mean-field approach used in \cref{sec:model_and_stationary,sec:homogeneous_dynamic}, we compare the time evolution of the trapped system obtained from TWA calculations to a mean-field simulation. For the latter, we set the initial virtual particle occupation to zero $\beta_n = 0$. The impurity velocity $v(t) = \Big(p(t) - \pB(t)\Big)/M$ and position $r(t)$ obtained in that way are shown in \cref{fig:trap_p_r}. For a small coupling constant $\gib \ll gn \xi$ the evolution remains sinusoidal, however, at large coupling it deviates strongly from an harmonic motion. The agreement between TWA and mean-field is reasonably well, given that we consider a quite strongly interacting gas with $\gamma = 0.1$. The impurity position \cref{fig:trap_p_r}b) gets in some cases an overall shift between TWA and mean-field, the motion is however qualitatively  similar.  
Especially the deviation in the impurity velocity is small, see \cref{fig:trap_p_r}a). From this, we conclude that the mean-field simulation is sufficient to predict the time evolution at least qualitatively.

\section{Summary}

In the present paper we have discussed the dynamics of the formation of a Bose polaron when an impurity is injected into a 1D weakly-interacting Bose gas by performing time-dependent simulations of mean-field equations of the condensate amplitude complemented by truncated Wigner simulations to include quantum fluctuations. In a homogeneous gas with periodic boundary conditions the total momentum $p$ of the system is conserved and can be used as independent parameter to characterize different dynamical regimes. Analyzing steady state solutions 
of the mean-field equations first, we showed that stationary solutions exist only for impurity velocities below a critical value, which in the limit of weak impurity-boson couplings $\gib$ agrees with the Landau critical velocity, as predicted by the Froehlich model, but monotonously decreases with increasing interaction and eventually approaches zero. This is because with growing values of $\gib$ the condensate is more and more depleted in the vicinity of the impurity, which leads to a reduced local speed of sound. While, as first shown in \cite{Hakim1997} for a given velocity of the impurity below the critical value, there are always two stationary solutions of the condensate equations, the solution is  unique when fixing the total momentum. For momentum values with a convex energy-momentum relation, $\partial^2 E/\partial p^2 >0 $, one of the two solutions applies and in regions with $\partial^2 E/\partial p^2 <0$ the other solution holds.
We showed moreover that the stationary energy-momentum relation is periodic in $p$ as the background condensate away from the impurity can pick
up additional quantized amounts of momentum corresponding to integer windings of the condensate phase over its length. 
As a consequence the relation between impurity velocity $v$ and polaron momentum $p$ is also periodic and includes regions of momentum where the impurity velocity is negative. In these regions the Bose gas stabilizes only a steady state with momentum exceeding the total momentum which must then be compensated by an opposite motion of the impurity. 
While a direct measurement of the energy-momentum relation of the polaron is challenging, its non-monotonous form can have interesting experimental consequences. E.g. injecting impurities with finite velocity into a small ring condensate can induce a finite circular current corresponding to a finite number of enclosed flux quanta.

To study the formation of the polaron we considered two cases, a slow, quasi-adiabatic turning on of the Bose-impurity coupling and a sudden quench. If in the quasi-adiabatic situation the initial velocity is chosen small enough such that it stays below the critical value at all times, the impurity is decelerated only due to the increase in its effective mass, associated with the formation of the polaron. As the system evolves quasi-adiabatic this reduction in velocity is  reversible. 
We find that the polaron quasiparticle is formed on the timescale of the inverse chemical potential $1/gn$ (see \cref{fig:velocity_quench}), which for parameters of a recent experiment in 1D gases \cite{Catani2012} is on the order of $60 \mathrm{\mu s}$.
If in the quasi-adiabatic scheme the initial velocity is above the critical value the impurity emits density waves irrespective how slowly the
interaction is turned on leading to irreversible friction. Switching on the impurity-boson interaction suddenly a rich scenario of dynamical regimes is observed. Depending on the mass ratio of particles, the total momentum and the impurity-boson coupling strength the impurity is slowed down by emission of density waves or grey solitons. The latter happens for large momenta and large impurity masses and is specific for the regime of strong impurity-boson coupling. In this case asymptotic states can form where the impurity velocity changes its sign, 
i.e. backscattering occurs, which cannot occur in the weak coupling regime dominated by Cherenkov radiation of phonons. While an in-situ measurement of the impurity motion is difficult in an experiment, the emission of grey solitons can be directly observed by density measurements of the condensate. We here considered impurities in one-dimensional condensates. The modified mean-field approach including the backaction of the impurity to the condensate can however be applied also to higher dimensions. Theories predicting the polaron dynamics based on the Froehlich model, using e.g. a coherent variational ansatzes \cite{Shchadilova2016,Drescher2019,Ardila2021} or master equation \cite{lausch2018,Nielsen2019} are capable of capturing the evolution as long as the condensate deformation is not substantial. From a straightforward dimensional analysis of the mean-field equation in $D$ dimension we estimate that the condensate deformation becomes significant for $ \gib / g \gtrsim n\xit^D $. Some of the predicted effects are expected to carry over from one to higher dimensions. E.g. the reversible slowdown of a sub-sonic impurity due to the formation of a polaron and the friction forces experienced by a super-sonic impurity due to emission of density waves will be very similar. The emission of grey solitons and the dragging of the impurity towards the grey solitons, on the other hand, is an effect specific to one-dimensional gases. In two dimensions a heavy, supersonic impurity might instead emit  vortex anti-vortex pairs.

To justify the validity of the mean-field approximation we performed truncated Wigner simulations of the full quantum problem in a trapped gas. The TWA accounts for quantum fluctuations due to Bogoliubov phonons on the deformed condensate background up to quadratic order. To avoid infrared divergencies related to the one-dimensional setup, enforced by the Mermin-Wagner-Hohenberg theorem, we considered a harmonically trapped gas. Although the total momentum is no longer conserved it follows a simple equation of motion, which we solve in semiclassical approximation. The TWA simulations show that the mean-field description of the dynamics of polaron formation is well justified as long as the Tonks parameter of the Bose gas is small, i.e. for a weakly interacting gas. The case of strong boson-boson interactions requires different analytical and numerical tools and will be discussed elsewhere.

\section*{Acknowledgement}
We would like to thank Artur Widera, Jonas Jager and Ryan Barnett for fruitful discussions. Financial support by the DFG through SFB/TR 185, Project No.277625399 is gratefully acknowledged. M.W. was supported by the Max Planck Graduate Center with the Johannes Gutenberg-Universität Mainz. 


\appendix

\section{Stationary mean-field solution}\label{sec:app_stationary_state}

\begin{figure*}[ht]
    \centering
    \includegraphics[width=\textwidth]{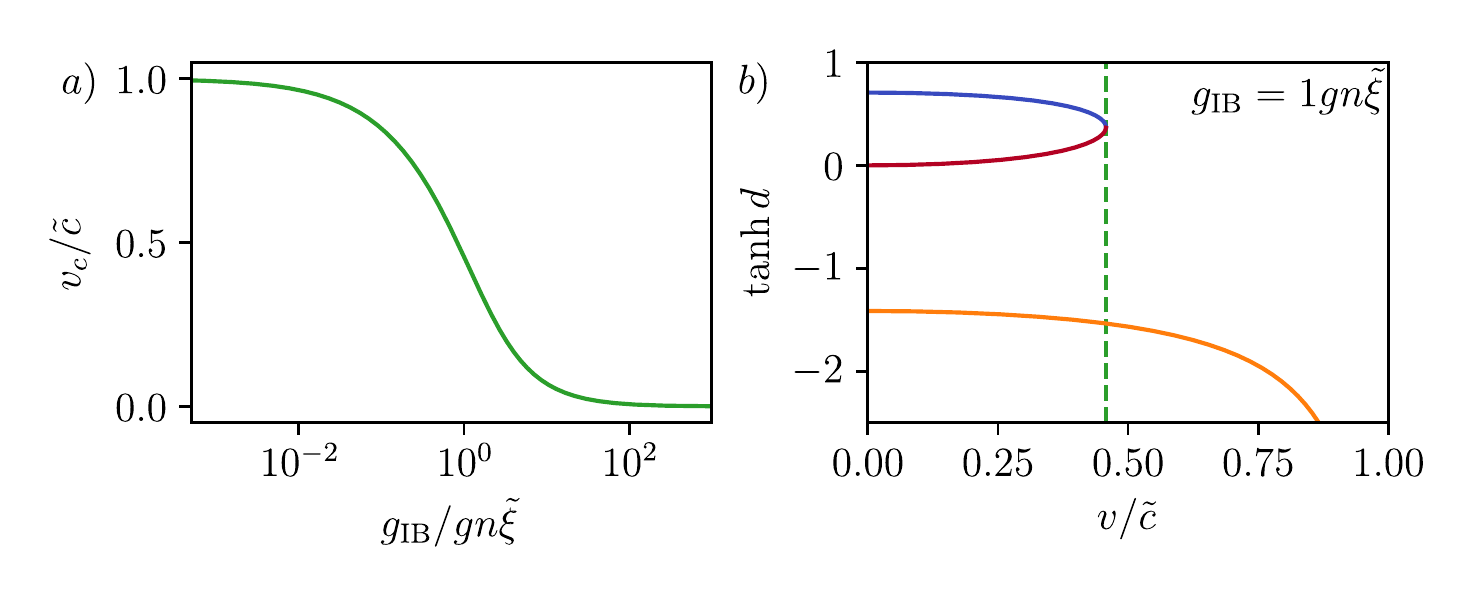}
    \caption{
    a) Critical impurity velocity $v_c$. b) Solutions of the cubic \cref{eq:tanhd}. The orange solution (lowest curve) is $\tanh{d}<-1$ for all $v$ and therefore nonphysical. The red and blue solutions exist only below the critical velocity $v_c$ (dashed green), yielding the two relevant states.
    }
    
    \label{fig:crit_momentum}
\end{figure*}

In the following, we briefly summarize the stationary solution of the GPE \cref{eq:GPE_homo}, which are derived in more detail in \cite{Hakim1997,Jager2020}. Since the equation explicitly depends on the impurity velocity $v$ and not the conserved momentum $p$, it is convenient to use $v$ as a parameter and then calculate the total momentum of the stationary state by
\begin{equation}
    p = M v + \pB  = M v  -i \int dx \phi^*(x,t) \partial_x \phi(x,t). \label{eq:app_momentum}
\end{equation}
This is possible since $v(t)$ is a constant in the steady state. As shown in \cite{Hakim1997,Jager2020} the stationary solution of \cref{eq:GPE_homo} is similar to a grey soliton, except at $x=0$, and given by
\begin{equation}
    \phi(x,t) = \sqrt{n} \, e^{i (\varphi_1 x +\varphi_2 \text{sgn}(x) - gn \, t)} \; \Big[a - i b \;\text{sgn}(x) \tanh \big(\frac{b}{\sqrt{2} \xit} |x| + d) \Big], \label{eq:stationary_state}
\end{equation}
as long as the system size $L$ is large compared to the rescaled healing length $\xit = 1/\sqrt{2 gn \tilde m} $. Here $a = v /  \tilde{c}$, $b = \sqrt{1-a^2}$ and the parameters $\varphi_1$ and $\varphi_2$ are chosen such that the phase of the solution is continuous at $x=0$ and fulfills periodic boundary condition. 
The parameter $d$ shifts the grey soliton wave function such that the boundary condition generated by the delta distribution in \cref{eq:GPE_homo}
\begin{equation}
    \partial_x \phi(x,t)\Big|_{x=0^-}^{0^+} = 2 \gib \tilde{m} \; \phi(0,t)
\end{equation}
is fulfilled. From this, it can be deduced that $\tanh{d}$ must be the solution of a cubic equation
\begin{equation}
    b^3 \tanh{d} \Big(1- \tanh^2{d} \Big)\overset{!}{=} \frac{\gib}{\sqrt{2} gn \xit} \Big(1-b^2+b^2 \tanh^2{d} \Big). \label{eq:tanhd}
\end{equation}
The three solutions of the equation are shown in \cref{fig:crit_momentum} b). The one which is real for all parameters is always less than or equal $-1$, such that $d$ is not a real number corresponding to a nonphysical state. The other two solutions are real and between $0$ and $1$, if the impurity velocity $v$ is below the critical velocity $v_c = a_c \,  \tilde{c}$, see \cref{fig:crit_momentum} a). Here $a_c$ can be determined by solving
\begin{equation}
    \frac{\gib}{g n \xit}  \overset{!}{=} \frac{1}{2 a_c} \sqrt{1-20a_c^2-8a_c^4 + (1+8a_c^2)^{3/2}},
\end{equation}
which is equivalent to a cubic equation in $a_c^2$. For a small coupling constant $\gib \ll gn \xit$ the critical velocity is $ \tilde{c}$ and agrees with the prediction of the Froehlich model \cite{Landau1941,lausch2018}. However for strong repulsion  $\gib \gg gn \xit$ the critical velocity converges to zero. 
Substituting the two physical solutions of \cref{eq:tanhd} into \cref{eq:stationary_state} yields the two stationary states mentioned in the main part of this work. The two solutions are equal at the critical momentum, explaining why the stationary states merge at criticality. 

Next, since $p$ and not $v$ is conserved under time evolution it is important to derive an expression relating the parameters for the stationary solutions. It follows from  \cref{eq:app_momentum,eq:stationary_state} and is given by
\begin{equation}
    p = M v - 2 \,n a b\, \big(1-\tanh{d}\big) 
    + 2 n  \Big[ \arctan{ \big( \frac{a}{b} \big) }  - \arctan{ \big( \frac{a}{b} \tanh{d} \big) } 
    \Big]. \label{eq:stationary_momentum}
\end{equation}
In order to compare this analytic expression to the time-dependent simulation we solve it numerically for $v$.

The polaron energy is given by $E_\text{pol} = E(\gib,p) - E(\gib=0,v=0)$, where $E(\gib,p)$ is the expectation value of the LLP Hamiltonian  \cref{eq:H_homo_LLP}
\begin{equation}
\begin{aligned}
    E_\text{pol} =& \;\sqrt{2} gn^2 \xit \Big[ 2 b \big(1- \tanh(d)\big)  - \tfrac{1}{3} b^3 \big(2-3 \tanh(d) + \tanh(d)^3\big) \Big] \\
                 &+ \half M v^2 \Big[1- 4 \sqrt{2} \, \frac{\tilde{m}}{M}\,  n \xit \big(1- \tanh(d)\big) \Big]. \label{eq:stationary_energy}
\end{aligned}
\end{equation}
In the approach described so far \cref{eq:stationary_momentum} results only in momentum values with $- \pi n\leq p \leq \pi n$. In order to reach higher momenta the stationary solution \cref{eq:stationary_state} must be modified by an additional phase gradient
\begin{equation}
    \tilde{\phi}(x,t) = e^{i x \frac{2 \pi}{L} \nu} \,  \phi(x,t) \quad \textrm{with} \; \nu \in 	 \mathbb{Z}.
\end{equation}
Except for the additional phase gradient, the stationary solution, all parameters in \cref{eq:stationary_state}, and the energy \cref{eq:stationary_energy} are not modified in the thermodynamic limit $L \gg \xit$.  Only the total momentum 
\begin{equation}
    \tilde{p} = p + 2 \pi n\, \nu
\end{equation}
picks up an additional term, which explains why the observables in \cref{fig:dispersion}a) and \cref{fig:mass}b) are periodic in $p$, with a period length of $2 \pi n$.

\section{Gapless adiabaticity}\label{sec:app_adiabatic2}

\begin{figure}%
    \centering
    \includegraphics[width=\textwidth]{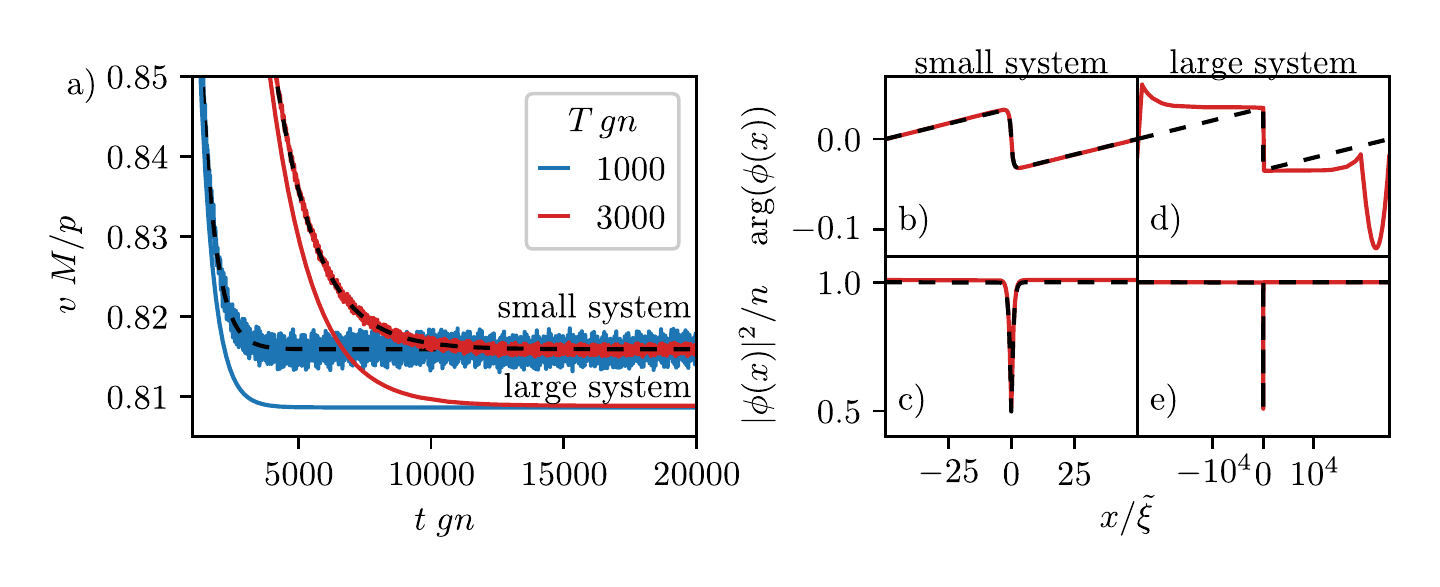}

    \caption{
    Comparison of the quasi-adiabatic (large system) to true-adiabatic (small system) evolution for $v(0) = p/M = 0.1 \tilde{c}$, $\gib = g n \xit$, $\gamma = 0.1$ and $M = 3m$.  a) Evolution of the impurity velocity for two turn-on timescales $T$. Dashed lines represent the instantaneous stationary state. b) - d) State at the end of the evolution at $t \,gn = 2 \cdot 10^4$ for $T\,gn =3 \cdot 10^3$. For a small system  $L=100 \tilde{\xi}$, both phase b) and density c) agree with the instantaneous stationary state (dashed).  For a large system $L =  5 \cdot 10^4 \xit$ the phase d) disagrees at a large distance from the center.      
    }
    \label{fig:adiabatic_system_size}%
\end{figure}

In \cref{sec:adiabatic} we showed that the system evolves quasi-adiabatic if the impurity-Bose coupling constant is turned on slowly compared to the other timescales. However, there always remains a small but finite difference to the instantaneous stationary state in \cref{fig:adiabatic}c). In this section we show, that this difference originates from the system not being energetically gaped in the thermodynamic limit. To this end, the time evolution of a large system $L \gg \tilde{c} T$ is compared to a small one $L \ll \tilde{c} T$, with otherwise equal parameters.
The small system is gaped due to finite-size effects, such that the adiabatic theorem strictly holds. This is shown in \cref{fig:adiabatic_system_size} a)-c), where the impurity momentum as well as the density and phase of the Bose gas agree with the instantaneous ground state. 
In contrast in the large system. Here in particular the phase disagrees at a large distance from the impurity $|x| \gg \xit$, see \cref{fig:adiabatic_system_size} d), explaining the small discrepancy of the impurity momentum \cref{fig:adiabatic_system_size}a). 
In the large system  the stationary state is not reached globally, but only locally at the position of the impurity. This is however sufficient for the system to evolve quasi-adiabatic.

\section{Bogoliubov-de Gennes in a trap}\label{sec:BdG}
In order to express the initial ground state of the trapped Bose gas, before the interaction with the impurity, we diagonalize the Bose gas Hamiltonian
\begin{equation}
    \hH_\text{B} = \int \dd x \, \hphid{x} \Big( - \frac{\partial_x^2}{2m} + \half m \omega^2 x^2   +\half g \hphid{x}  \hphi{x} \Big)  \hphi{x}, \label{eq:H_trapped_Gas}
\end{equation}
approximately using a Bogoliubov-de Gennes (BdG) approach \cite{Pethick2008}. In the first step, the mean-field ground state is determined by the Gross-Pitaevskii equation (GPE)
\begin{equation}
    \Big( - \frac{\partial_x^2}{2m} + \half m \omega^2 x^2 + g |\phi_0(x)|^2 - \mu \Big) \phi_0(x) = 0,
\end{equation}
which we solve numerically using imaginary time evolution. Here $\mu$ is the mean-field chemical potential. In case of a weakly interacting Bose gas, it is sufficient to only include small fluctuation on top of the mean-field solution, which is done by expressing the bosonic field operators by
\begin{equation}
    \hphi{x} = \phi_0(x) + \sum_n \Big(u_n(x)  \hat{b}_n + v_n(x)^* \hat{b}^\dagger_n \Big),
\end{equation}
and only keep terms up to quadratic order in the operators $\hat{b}_n^{(\dagger)}$. Here $u_n(x)$ and $v_n(x)$ are the BdG coefficients. This Ansatz diagonalizes the Hamiltonian \cref{eq:H_trapped_Gas} if the coefficients fulfill the BdG equation
\begin{equation}
\begin{aligned}
    &\left(
        \begin{matrix}
        \hat{L} & g |\phi_0(x)|^2 \\
        -g |\phi_0(x)|^2 & -\hat{L}
        \end{matrix}
    \right)
    \left(
        \begin{matrix}
        u_n(x)  \\
        v_n(x)
        \end{matrix}
    \right)
    = 
    \epsilon_n  
    \left(
        \begin{matrix}
        u_n(x)  \\
        v_n(x)
        \end{matrix}
    \right),\\
    &\textrm{where} \qquad \hat{L} =- \frac{\partial_x^2}{2m} + \half m \omega^2 x^2  + 2 g |\phi_0(x)|^2 - \mu,
\end{aligned}
\end{equation}
where $\epsilon_n$ are the eigenenergies of the corresponding BdG modes. To solve this equation, we expand it in a finite number of eigenfunctions of the free harmonic oscillator and diagonalize the resulting matrix numerically.


\section*{References}
\bibliography{library.bib}
\end{document}